Article

# Enhanced Algal Photosynthetic Photon Efficiency by Pulsed Light

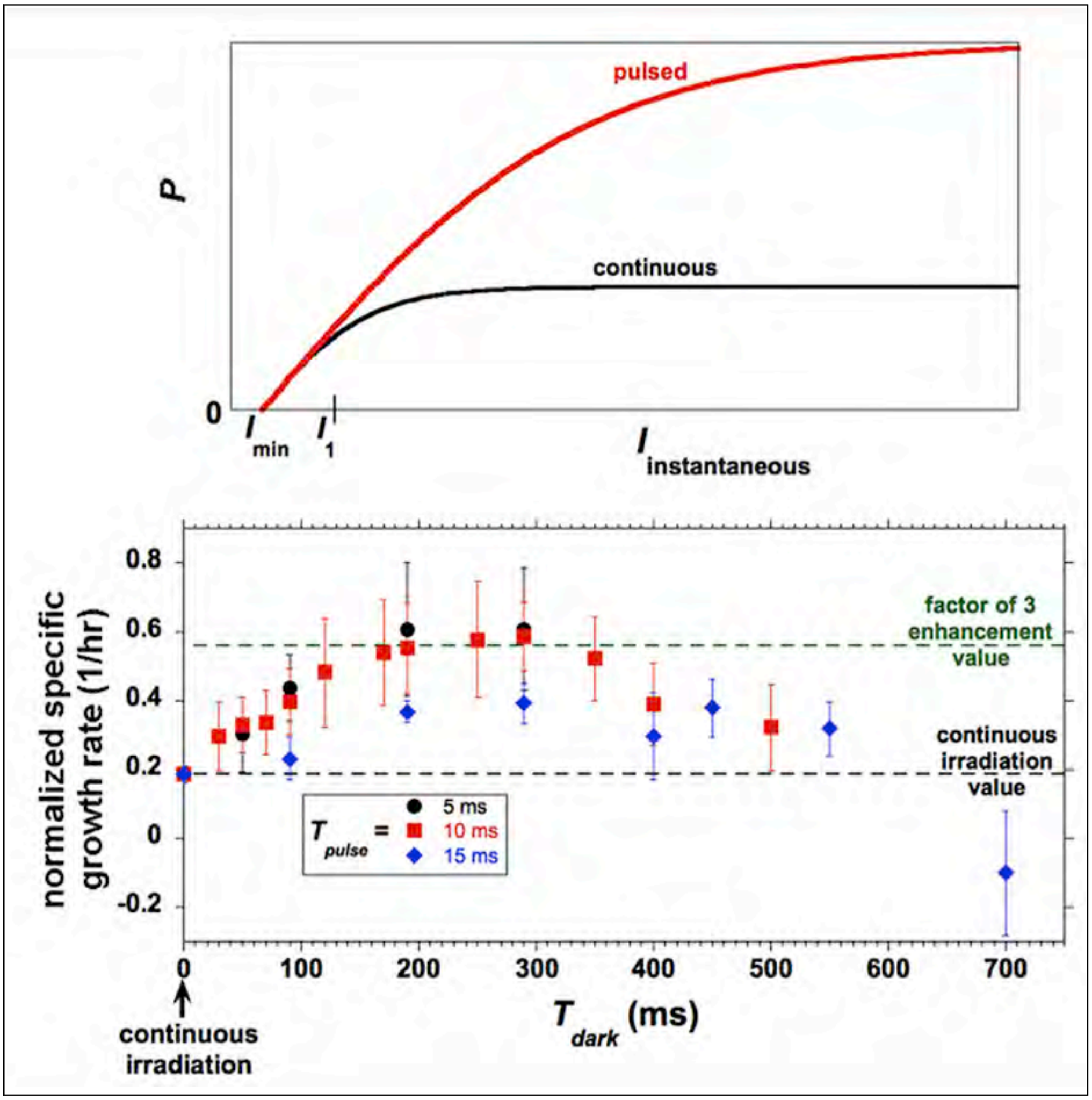

Yair Zarmi, Jeffrey M. Gordon, Amit Mahulkar, Avinash R. Khopkar, Smita D. Patil, Arun Banerjee, Badari Gade Reddy, Thomas P. Griffin, Ajit Sapre

zarmi@bgu.ac.il

## HIGHLIGHTS

Sizable enhancement of algal photosynthetic photon efficiency at high light intensity

Extensive experimental evidence from new and revisited experiments

Model based on photon arrival statistics accounts for all experimental observations

The key is synchronizing biological and photonic timescales via pulsed light

Article

# Enhanced Algal Photosynthetic Photon Efficiency by Pulsed Light

Yair Zarmi,[1,5,*] Jeffrey M. Gordon,[1,2] Amit Mahulkar,[3] Avinash R. Khopkar,[3] Smita D. Patil,[3] Arun Banerjee,[3] Badari Gade Reddy,[3] Thomas P. Griffin,[3,4] and Ajit Sapre[3]

## SUMMARY

**We present experimental results demonstrating that, relative to continuous illumination, an increase of a factor of 3–10 in the photon efficiency of algal photosynthesis is attainable via the judicious application of pulsed light for light intensities of practical interest (e.g., average-to-peak solar irradiance). We also propose a simple model that can account for all the measurements. The model (1) reflects the essential rate-limiting elements in bioproductivity, (2) incorporates the impact of photon arrival-time statistics, and (3) accounts for how the enhancement in photon efficiency depends on the timescales of light pulsing and photon flux density. The key is avoiding "clogging" of the photosynthetic pathway by properly timing the light-dark cycles experienced by algal cells. We show how this can be realized with pulsed light sources, or by producing pulsed-light effects from continuous illumination via turbulent mixing in dense algal cultures in thin photobioreactors.**

## INTRODUCTION

### Background and Motivation

The quest to improve the bioproductivity of algae is largely prompted by their value for biofuels, nutritional supplements, and pharmaceuticals (Borowitzka and Moheimani, 2013). Previous investigations have focused on (1) trying to improve the efficiency with which the photosynthetic apparatus can exploit incident photons as well as (2) more efficiently distributing those photons among the algae in a photobioreactor (Tennessen et al., 1995; Gebremariam and Zarmi, 2012; Greenwald et al., 2012; Zarmi et al., 2013; Gordon and Polle, 2007; Abu-Ghosh et al., 2015).

Algae exhibit maximal conversion efficiency at photon flux densities $I$ that are below ~150–300 μE/(m$^2$-s), denoted by $I_1$ in Figure 1 (1 E ≡ 1 mole of photons, and only photosynthetically active radiation is considered, i.e., wavelengths from 400 to 700 nm). However, high rates of biomass production require maintaining maximum efficiency at the far higher light intensities characteristic of peak solar radiation (~2,000 μE/(m$^2$-s)), which is not achieved by cultivating algae under continuous light, where a progressively increasing fraction of the photons above intensity $I_1$ are dissipated as $I$ increases. A pivotal question, then, is to what extent this inefficiency can be surmounted by illumination with pulsed (rather than continuous) light.

The consensus from an extensive literature on algal photobioreactor performance has been that pulsed light cannot lead to higher bioproductivity (Schulze et al., 2020; Graham et al., 2017). In these studies, however, the biomass production rate was taken as the average over the entire cycle time, i.e., over both light and dark periods. Results from pulsed-light experiments were compared against those under continuous light of the same *cycle-average* photon flux density, and not the instantaneous intensity of the pulsed light, such that the number of photons impinging on the culture over an extended period of time was the same for both the pulsed and continuous light protocols.

Comparisons were sometimes generated for continuous-light intensities $I < I_1$, where maximal efficiency is achieved, so, intrinsically, no improvement was possible with pulsed light. Other studies used cycle and pulse times that are well beyond the values where efficiency enhancements are possible (Schulze et al., 2020; Graham et al., 2017; Combe et al., 2015). And publications depicting pulsed-light experiments where enhancements in photosynthetic efficiency could have been deduced are sparse and do not explicitly note the improvements (Vejrazka et al., 2011, 2012, 2013, 2015; Simionato et al., 2013).

[1]Department of Solar Energy and Environmental Physics, Jacob Blaustein Institutes for Desert Research, Ben-Gurion University of the Negev, Sede Boqer Campus 8499000, Israel

[2]School of Mechanical and Chemical Engineering, University of Western Australia, Perth WA, 6009, Australia

[3]Reliance Industries Ltd., Mumbai, MH, India

[4]Present address: Breakthrough Energy Ventures, Boston, MA, USA

[5]Lead Contact

*Correspondence: zarmi@bgu.ac.il

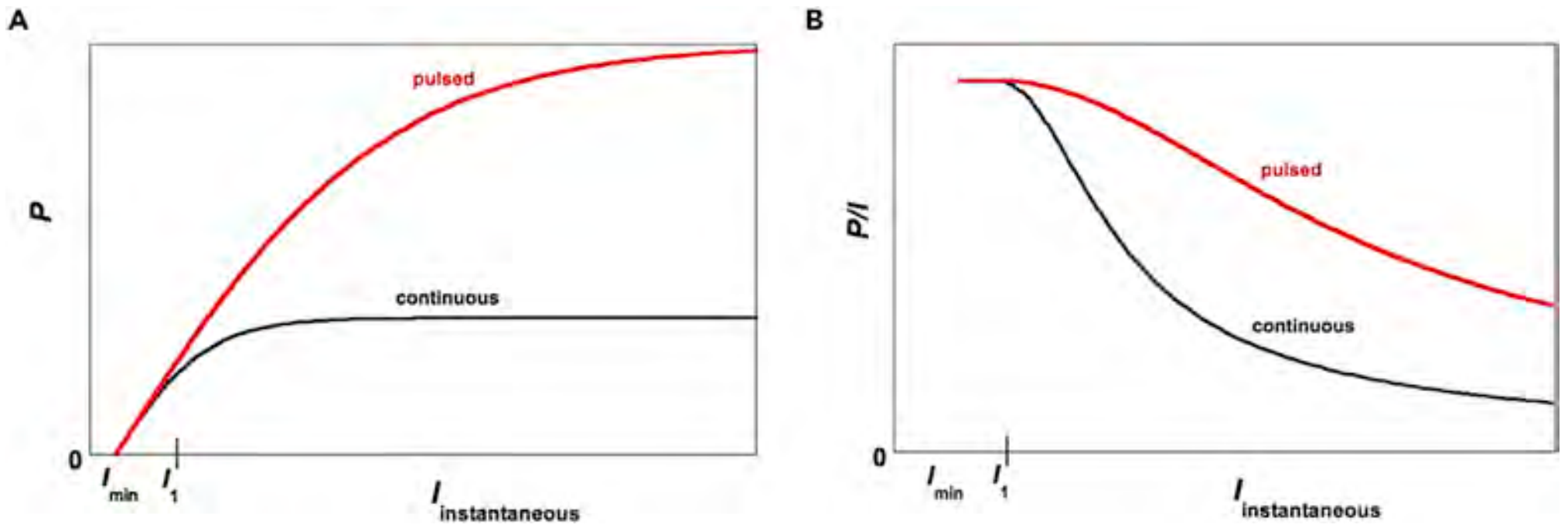

**Figure 1. Key Trends in Algal Photobioreactor Performance**

(A) Biomass production rate per unit time and per unit area, *P*, as a function of *instantaneous* photon flux density. Under continuous light, the instantaneous and time-averaged photon flux densities are identical. For pulsed light, however, $I_{instantaneous}$ refers to its value only during $T_{pulse}$, which is why *P* can be noticeably greater under pulsed light.

(B) The corresponding variation of *P/I* (biomass production *per photon*) with *instantaneous* photon flux density.

The purpose of this article is to show how such previous misconceptions can be resolved and, in the process, to present and analyze extensive experimental measurements showing how algal photosynthetic efficiency *per photon*, in particular at the light intensities required for high biomass generation rates, can be enhanced severalfold relative to continuous illumination via the judicious application of pulsed light.

## Basic Trends in Bioproductivity (*P-I* Curves)

The performance of algal photobioreactors is commonly plotted as bioproductivity *P* versus *I* (Figure 1). *P* may be the dry biomass production rate (e.g., g dry wt./($m^2$-s)) (Mann and Myers, 1968; Molina-Grima et al., 2000; Chisti, 2007), or the $O_2$ production rate (e.g., mol/($m^2$-s)) (Chalker et al., 1993; Geider and Osborne, 1992). Once *I* exceeds $I_{min}$ below which algal respiration dominates (the light compensation point), *P* increases linearly until, over a range of flux densities above $I_1$, *P* grows sublinearly and then plateaus (Figure 1A). Bioproductivity *per photon* is *P/I* (Figure 1B). As $I_{min}$ is invariably far smaller than *I* values of practical interest, *P/I* remains essentially constant at its maximum value up to around $I_1$, and then decreases with *I*.

For pulsed light, each cycle has an irradiation duration $T_{pulse}$ and a dark period $T_{dark}$. In the presentation that follows, *I* consistently refers to the *instantaneous* photon flux density, be it for continuous or pulsed light. To evaluate *P per photon*, one must take into account the number of photons (per unit area) that hit the photosynthetic apparatus during a pulse ($I \cdot T_{pulse}$). Hence, for comparisons with continuous light data, *P/I* must be computed only for the time of exposure to light pulses, rather than the full cycle time.

## Figure of Merit: Relative Photon Efficiency

To compare between pulsed and continuous light operation, we define relative photon efficiency $\eta_{ph}$ as:

$$\eta_{ph} = \left(\begin{array}{c}\text{Average biomass generated per photon}\\ \text{in pulsed regime}\end{array}\right) \Big/ \left(\begin{array}{c}\text{Average biomass generated per photon}\\ \text{under continuous light operation}\end{array}\right). \quad \text{(Equation 1a)}$$

Assuming the number of photons reaching the reaction centers is proportional to $T_{pulse}$, we re-express $\eta_{ph}$ as:

$$\eta_{ph} = \left(\frac{\text{Average biomass generated in one cycle}}{T_{pulse}}\right) \Big/ \left(\begin{array}{c}\text{Average biomass production rate}\\ \text{under continuous illumination}\end{array}\right). \quad \text{(Equation 1b)}$$

Equation 1b constitutes a good approximation for $\eta_{ph}$ (in the statistical sense) as long as the number of photons hitting a reaction center during $T_{pulse}$ is sufficiently longer than the average time between the arrival of consecutive photons. For example, for $I$ = 1,000 μE/($m^2$-s) and an effective photon absorption cross-section *A* of 1 $nm^2$, the average time between photon arrivals is 1.66 ms. *Effective* absorption cross-section refers to a characteristic value for each reaction-center antenna responsible for photochemical conversion in Photosystem II (PS II). Extensive measurements have established a value of

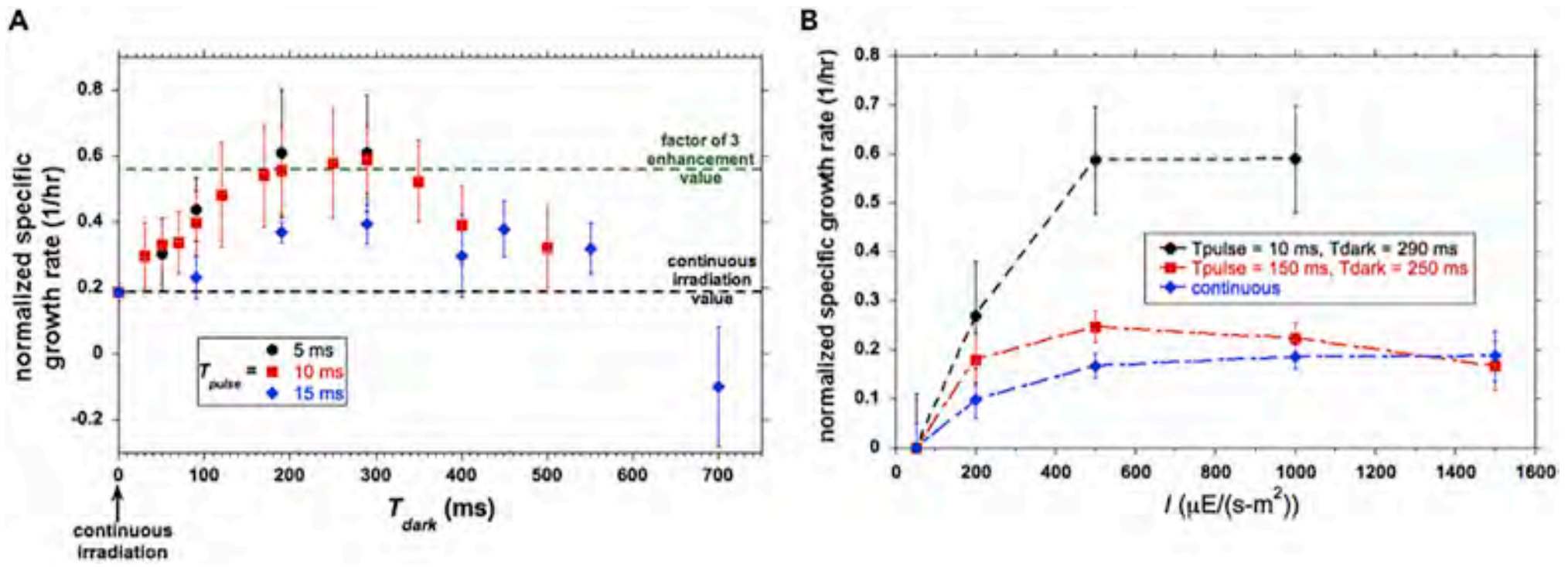

**Figure 2. Measured Normalized Specific Growth Rate**

As a function of: (A) $T_{dark}$ for pulsed light with $I$ = 1,000 μE/(m$^2$-s) at $T_{pulse}$ = 5, 10, and 15 ms and (B) I for both continuous irradiation and pulsed light. Vertical bars indicate ±1 standard deviation about the average. Each data point comes from 20 replications for experiments with continuous light and 8 replications for experiments with pulsed light.

the order of 1 nm$^2$, which, depending on algal strain and photo-acclimation, can vary from about half that value up to several square nanometers (de Wijn and van Gorkom, 2001; Zou and Richmond, 2000; Simionato et al., 2011; Bonente et al., 2012; Gris et al., 2014; Ley and Mauzerall, 1982; Klughammer and Schreiber, 2015; Osmond et al., 2017; Murphy et al., 2017; Koblízek et al., 2001). In contrast, the measured cross-section of a single chlorophyll molecule is ~0.003 nm$^2$ (Ley and Mauzerall, 1982). However, the effective absorption cross-section $A$ for each reaction center antenna, which comprises of the order of hundreds of chlorophyll molecules, is commensurately larger (Ley and Mauzerall, 1982; Klughammer and Schreiber, 2015; Osmond et al., 2017; Murphy et al., 2017; Koblízek et al., 2001; Greenbaum, 1988).

Exploiting the statistics of photon arrival times, we will show below that, for $I$ = 1,000 μE/(m$^2$-s) and $A$ = 1 nm$^2$, $T_{pulse}$ should be longer than ~5 ms, and that at $I$ = 250 μE/(m$^2$-s), $T_{pulse}$ should be longer than ~10 ms. If, on the other hand, $T_{pulse}$ is short relative to the average photon arrival time, then Equation 1b is not a suitable expression for $\eta_{ph}$ and must be revised. *Absolute* photon efficiency can vary noticeably with reactor temperature, reactor chemistry, algal strain, and light intensity. However, the primary aim here is demonstrating how properly chosen pulsed-light protocols can boost photon efficiency relative to continuous illumination. Hence relating to *relative* photon efficiency permits a meaningful comparison among previous studies and against the measurements reported below.

### Deducing Photon Efficiency Enhancements from Prior Studies

Published data for pulsed-light experiments where it is possible to rigorously deduce enhancements in $\eta_{ph}$ are scarce, and do not explicitly articulate them. The publications for which the reported data permit one to ascertain such improvements are: (Vejrazka et al., 2011, 2012, 2013, 2015; Simionato et al., 2013).

From the data in Vejrazka et al. (2011, 2012, 2013, 2015), $\eta_{ph} \approx 3$ can be deduced for $T_{pulse}$ = 1 ms and $T_{dark}$ = 9 ms at $I$ = 1,000 μE/(m$^2$-s). From the data in Simionato et al. (2013), $\eta_{ph} \approx 3$ can be determined for $T_{pulse}$ = 11 ms and $T_{dark}$ = 22 ms at $I$ = 350 μE/(m$^2$-s), with $\eta_{ph} \approx 10$ for $T_{pulse}$ = 10 ms and $T_{dark}$ = 90 ms at $I$ = 1,200 μE/(m$^2$-s). Both are elaborated in the following discussion.

## RESULTS

### New Experimental Measurements of Marked Increases in $\eta_{ph}$ under Pulsed Light

Our experimental methods are summarized in Transparent Methods of the Supplemental Information, including the definition of specific growth rate $\mu$. The normalized $\mu$ plotted in Figure 2 is defined so as to permit comparisons between pulsed and continuous illumination experiments, equal to the product of (1) average biomass production rate over a long time and (2) ($T_{pulse}$ + $T_{dark}$)/$T_{pulse}$. Hence the normalized $\mu$ is the average biomass produced in one cycle divided by $T_{pulse}$.

Notable aspects of our results are summarized in Figure 2. Figure 2A highlights that at $I$ = 1,000 μE/(m$^2$-s): (1) $\eta_{ph}$ is enhanced by a factor of ~3, (2) $\eta_{ph}$ increases and peaks as $T_{dark}$ lengthens, (3) an optimal $T_{dark}$ exists beyond which $\eta_{ph}$ decreases, and (4) the highest $\eta_{ph}$ is achieved for shorter pulses and lessens as $T_{pulse}$ is increased beyond ~10 ms. Figure 2B shows the dependence of $\eta_{ph}$ on $I$.

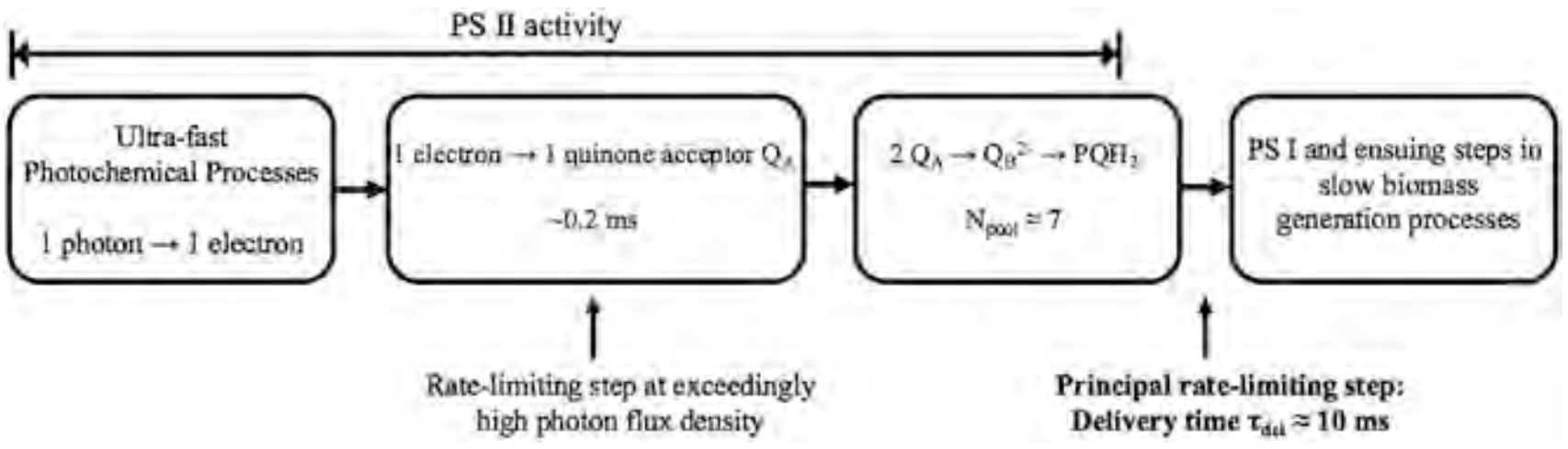

**Figure 3. Schematic of the Rate-Limiting Step in Algal Bioproductivity as Implemented in Our Model**

The existence of an optimal dark time may be understood by considering extreme cases. (1) For very short $T_{dark}$, there is not enough time to process all plastoquinones (*PQ*s) the reduction of which forms an essential link in the photosynthetic chain, including the *PQ* pool, which refers to whether the *PQ* molecules exist in an oxidized or reduced state (see Figure 3). In this limit, $\eta_{ph}$ is close to its value under continuous illumination. (2) For excessive $T_{dark}$, *PQ*s may decay or the system may revert to respiration.

Our highest $\eta_{ph}$ was achieved for $T_{dark}$ = 200–300 ms. The measurements of Simionato et al. (2013) spanned $T_{dark}$ = 22–900 ms. In the experiments of Vejrazka et al. (2011, 2012, 2013, 2015), much shorter pulses (1 ms) and dark times (9 ms) were applied. The magnitude of the average $T_{dark}$ in the thinnest and most efficient turbulent dense-culture flat-plate photobioreactors reported in Qiang et al. (1998a, 1998b) and Richmond et al. (2003) was 200–400 ms. In no prior investigation did the authors search for an optimal $T_{dark}$. We assume in our model that there is a dark time over which all *PQ*s reduced during a pulse are exploited for biomass production.

## The Model and the Rate-Limiting Process

Our model for the rate-limiting step of PS II, sandwiched between ultra-fast photochemical processes and far longer chemical processing in Photosystem I (PS I), is sketched in Figure 3. Biomass generation is treated as proportional to the number of charges delivered to PS I by re-oxidation of *PQ*s at Cytb$_6$f and subsequent production processes, with two photons being required for the reduction of one *PQ*.

The model allows for the confluence of three factors: (1) the potential to avoid "clogging" of the photosynthetic pathway inherent to continuous irradiation by proper timing of the light-dark cycles, (2) the state of the *PQ* pool, and (3) photon arrival-time statistics. We proceed by presenting the ideas incorporated in the model, followed by demonstrating that the results of Vejrazka et al. (2011, 2012, 2013, 2015) and Simionato et al. (2013) and our new data can all be accounted for by the model with reasonable values for the parameters $A$, *PQ* pool size $N_{pool}$, and delivery time $\tau_{del}$.

## "Bottleneck" Timescale and *P-I* Curves

Experiments indicate that the transition from linearity to saturation in algal *P-I* curves ($I = I_1$ in Figure 1) occurs in the range $I$ = 150–300 μE/(m$^2$-s). Our measurements plotted in Figure 4 below provide one such example. This transition can be related to timescales of physiological relevance. For example, for $A$ = 1 nm$^2$, a photon is then absorbed, on average, every 10 to 5 ms. So with two photons being required to reduce one *PQ* (Figure 3), an average of 20 to 10 ms is needed. If $A$ is doubled, then the requisite time is halved to 10 to 5 ms. Thus, the transition to saturation occurs when the average *PQ* reduction time is of the order of 10 ms. At higher $I$ (the saturation branch of the *P-I* curve) the average *PQ* reduction rate is clearly faster.

Saturation suggests there is a timescale of the same order of magnitude in the subsequent stage of the production process, which constitutes a "bottleneck." Indeed, it has been recognized that once a *PQ* is reduced, its delivery time $\tau_{del}$ to the next processing stage is ~10 ms (Diner and Mauzerall, 1973; Mauzerall and Greenbaum, 1989; Joliot and Joliot, 2008; Govindjee et al., 2010; Hasan and Carmer, 2012; Greenbaum, 1979), so that the average charge delivery rate to Cytb$_6$f (from PSII) on the saturation branch of the *P-I* curve should be 1/$\tau_{del}$.

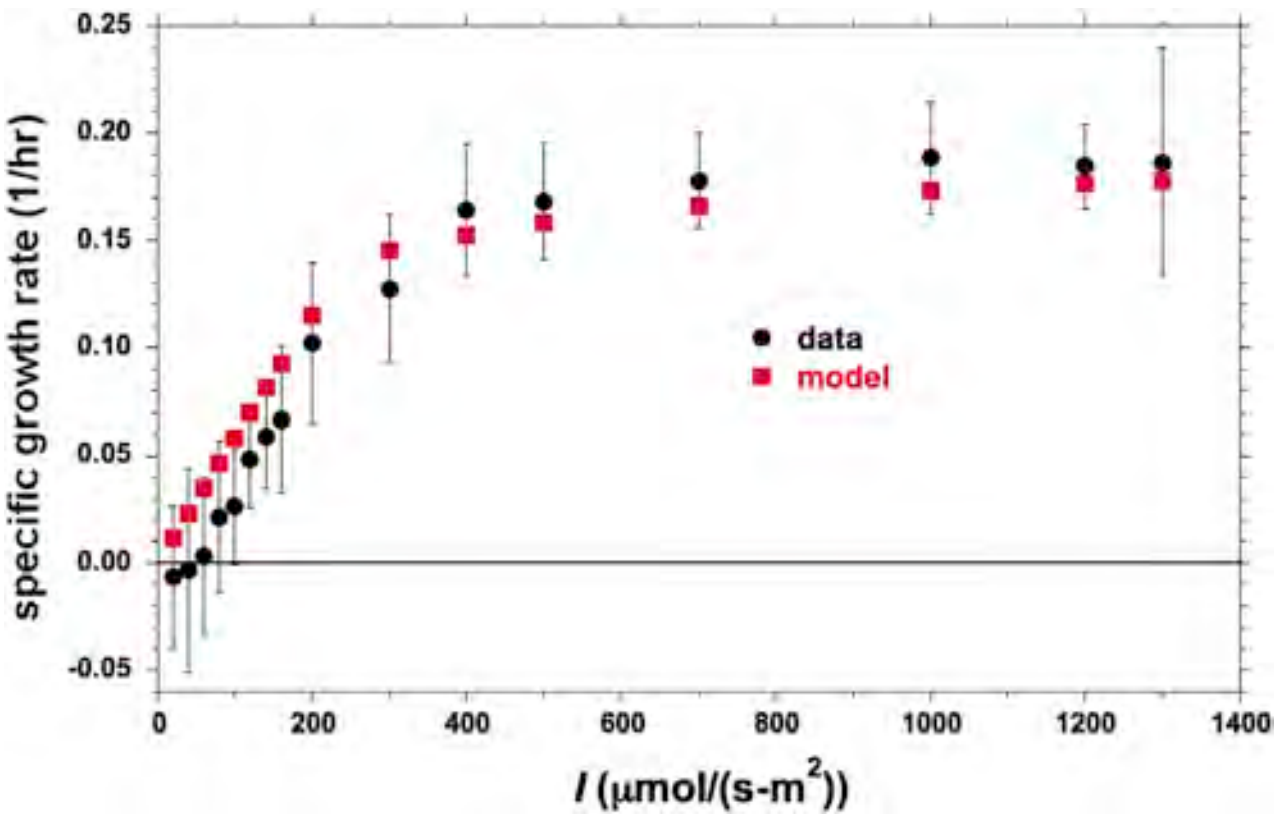

**Figure 4. Comparison of Model Predictions for a *P-I* Curve against Our Measurements for Continuous Illumination**
Vertical bars indicate $\pm 1$ standard deviation about the average for the measured data, with each point coming from 20 replications.

Ensuing production stages have processes with timescales at least one order of magnitude greater ($\geq$100–200 ms). However, they do not seem to affect *P*, based on the argument that if they gave rise to a bottleneck effect, then the transition to saturation in the *P-I* curve would occur at roughly 10 times lower values of *I*. In fact, the same bottleneck effect characterizes the rates of $O_2$ generation (a direct product of PS II) and biomass generation (a product of processes beyond PS II).

In view of the fact that the only observed bottleneck is a timescale related to the rate of reduction of *PQ*s, we adopt the hypothesis that the biomass production rate is proportional to the rate of reduction of *PQ*s by PS II to the following production stages.

## *PQ* Pool and Pulsed-Light Operation

Prior studies have estimated the size of the *PQ* pool in PS II as $N_{pool} \approx$ 5–12 with average values of $\sim$7, compounded by evidence that there may be a distribution of $N_{pool}$ (Simionato et al., 2013; Greenbaum, 1979; McCauley and Melis, 1986; Guemther et al., 1988; Hemelrijk and van Gorkom, 1996; Cleland, 1998). This pool enables PS II to store the energy extracted from the dissociation of up to $N_{pool}$ water molecules (driven by the absorption of up to $2 \cdot N_{pool}$ photons) by twice reducing each available *PQ*.

To appreciate the influence of $N_{pool}$ on $\eta_{ph}$, we first consider continuous irradiation. Along the linear part of a *P-I* curve in Figure 1A, each electron from *PQ* is delivered to the next electron carrier, so that *PQ* pool saturation does not impact performance. Once $I > I_1$, the *PQ* reduction rate exceeds the electron delivery rate. Hence, almost immediately after irradiation begins, the *PQ* pool is saturated, and excess photons are not utilized as all the charge carriers are occupied. Because *PQ* reduction rate increases with *I*, whereas the delivery rate is approximately independent of *I*, photon efficiency decreases with *I*.

Now consider pulsed operation at $I_{instantaneous} > I_1$ (Figure 1). During the pulse, some reduced *PQ*s may deliver charges to the subsequent stage. Concurrently, additional *PQ*s get reduced in the pool, because the *PQ* re-oxidation rate is slower than the reduction rate. If the light pulse ends just when the *PQ* pool has been reduced, namely, before "clogging" begins, and if the system is then given enough dark time to process all the *PQ*s reduced during the pulse and/or in the pool, then all absorbed photons can be exploited, and a correspondingly higher $\eta_{ph}$ can be attained (modulo losses that may occur during $T_{dark}$).

There is direct experimental evidence, not just correlative evidence, supporting the assumption about how the state of the PQ pool and its saturation affect photosynthetic dynamics (Joliot and Joliot, 1984a, 1984b, 1992; Joliot, 2003; Rokke et al., 2017; Suslichenko and Tikhonov, 2019). There is comparable additional evidence for the influence of the PQ pool on photosynthetic efficiency from investigations of direct hydrogen production (rather than biomass generation) from algae (Greenbaum, 1979) where the same rate-limiting steps in PS II dominate photosynthetic yield.

### Photon Arrival-Time Statistics

Photon arrival-time statistics are governed by a Poisson distribution:

$$P(\Delta t) = \frac{1}{\tau} e^{-\Delta t/\tau}, \quad \text{(Equation 2)}$$

where $\Delta t$ is the time between the arrival of two consecutive photons and $\tau$ is the average of $\Delta t$, which for the Poisson distribution is equal to the standard deviation. Note that this is *not* a narrow distribution centered on its average with a small standard deviation.

For example, consider $A = 1$ nm$^2$. At $I = 2{,}000$ μE/(m$^2$-s), 1,204 photons/s are absorbed on average, so the average time gap between two consecutive photons is 0.83 ms, with a standard deviation of 0.83 ms. Except for the effect of the 0.2 ms timescale, photon arrival statistics do not affect photon efficiency under continuous light, because averaging over long times depends only on the average photon arrival rate, to which $I$ is proportional. However, once light pulses as short as several milliseconds are considered, the statistics of photon arrival times plays a progressively important role in the *PQ* reduction rate.

### The 0.2-ms Timescale and Its Influence at High Light Intensity or Short Pulse Length

The excitation of a reaction center by a single photon, leading to the reduction of a *QA*, is estimated to require ~0.2 ms (Tietz et al., 2015). While the reaction center is engaged reducing a *QA*, it cannot process additional photons. Hence, if a second photon hits the reaction center within less than 0.2 ms, it is unutilized. This timescale plays a significant role at high photon flux densities (≥2,000 μE/(m$^2$-s)), as well as at lower flux densities under short pulses of the order of 1 ms.

Under continuous light, the loss of a second photon in a consecutive pair, which arrives within less than 0.2 ms after the first photon, is small compared with the losses owing to "clogging." For example, with $I = 2{,}000$ μE/(m$^2$/s) and $A = 1$ nm$^2$, the fraction of lost second photons is 11%, whereas the loss owing to the slowness of *PQ* re-oxidation is ~80%. If, on the other hand, a pulse duration that ensures no photons are lost due to "clogging" is applied, then the loss owing to the effect of the 0.2-ms timescale may constitute a significant component in the reduction of photon efficiency. In a 10-ms pulse, the loss is 11% of the photons. If *A* or *I* is doubled, then the loss amounts to ~21%. For a 1-ms pulse, this fraction increases to ~25%, for which doubling *A* or *I* increases the loss to 31%.

### Statistics of *PQ*s under Pulsed-Light Operation

The aforementioned observations, compounded with the experimental findings of de Wijn and van Gorkom (2001), Zou and Richmond (2000), Simionato et al. (2011), Bonente et al. (2012), Gris et al. (2014), Ley and Mauzerall (1982), Klughammer and Schreiber (2015), Osmond et al. (2017), Murphy et al. (2017), and Koblízek et al. (2001) lead us to propose that biomass generation is proportional to the number of *PQ*s delivered to the ensuing production processes. $\eta_{ph}$ calculated from the model is then

$$\eta_{ph} = \frac{\left(\dfrac{\text{Average number of } PQ\text{s reoxidized at cyt}b_6\text{f in one flash}}{T_{pulse}}\right)}{\left(\begin{array}{c}\text{Average reoxidation rate of } PQ \text{ at cytb}_6\text{f}\\ \text{under continuous illumination}\end{array}\right)} = \left(\frac{\text{Average number of } PQ\text{s reoxidized at cyt}b_6\text{f in one flash}}{T_{pulse}}\right) \bullet \tau_{del} \quad \text{(Equation 3)}$$

Owing to the statistical nature of photon arrival times, the number of *PQ*s re-oxidized under pulsed-light operation is a random variable with an average and a standard deviation. The average varies in a non-trivial manner with $T_{pulse}$ and the reduced state of the *PQ* pool. The standard deviation is large for short pulses and diminishes as $T_{pulse}$ is increased. When a distribution of *PQ* pool sizes is incorporated, as $T_{pulse}$ is increased, pools with a larger number of *PQ*s allow for the reduction of a larger number of *PQ*s and their subsequent re-oxidation during $T_{pulse}$. The effect of this possibility is demonstrated in the following discussion. Under continuous light, averaging over long exposure times smoothes out the fluctuations owing to photon statistics. By contrast, under short pulses, the measurements have significant statistical fluctuations, especially in measurements that do not extend over a long time.

## Confirming Model Validity for Continuous-Illumination Performance

Before embarking upon comparisons of model predictions against data from pulsed-light experiments, we must ensure that the model can predict *P-I* curves under continuous illumination. Our measured *P-I* curve is presented with our model calculation in Figure 4. The detailed computational procedure is described later in the discussion. The values of the biological parameters chosen were: $A$ = 1 nm$^2$, $N_{pool}$ = 7, and $\tau_{del}$ = 10 ms. The small effect of respiration at very low $I$ was not incorporated considering the error bars on the data, i.e., the light compensation point was taken as $I_{min} \approx 0$.

The model generates the rate of *PQ* re-oxidation at Cytb$_6$f. Assuming that the biomass production rate is proportional to the latter, the experimental and computed curves ought to be proportional to one another:

$$P_{Biomass} = C \bullet (\text{PQ reoxidation rate}) \qquad \text{(Equation 4)}$$

with a least-squares fit yielding $C$ = 1.956.

## Simple Average Model

### *Description*

Assuming a single value for $N_{Pool}$, we approximate $N_g$, the number of *PQ*s reduced per pulse, by

$$N_g = n_0 T_{pulse}/2. \qquad \text{(Equation 5)}$$

Here, $n_0$ is the rate of photons hitting the reaction center (photons/ms), given by

$$n_0 = \left(I \cdot 10^{-6} \cdot Av\right)\left(A \cdot 10^{-18}\right), \qquad \text{(Equation 6)}$$

where $I$ is in $\mu$E/(m$^2$-s), $Av$ is Avogadro's number (6.02×10$^{23}$), $A$ is in nm$^2$, and $T_{pulse}$ is in ms. When $T_{pulse}$ exceeds the *PQ* reduction time, the number of *PQ*s reduced during the pulse is approximately

$$n_o^{re-oxidized} = T_{pulse}/\tau_{del}. \qquad \text{(Equation 7)}$$

The number of *PQ*s reduced during the pulse and not re-oxidized immediately is then:

$$n_{additional} = N_g - n_o^{re-oxidized} = \left(\frac{n_o}{2} - \frac{1}{\tau_{del}}\right) T_{pulse} \qquad \text{(Equation 8)}$$

which are stored in the *PQ* pool as long as it is not full. The number of additional (reduced) *PQ*s should not exceed the maximum allowed pool size:

$$n_{additional} \le N_{Pool} \quad \Rightarrow \quad T_{pulse} \le \frac{N_{Pool}}{(n_0/2 \; - \; 1/\tau_{del})} \equiv T_{pulse}^{Max}. \qquad \text{(Equation 9)}$$

If $T_{pulse}$ does not obey Equation 9, then the *PQ* pool is completely reduced at the end of the pulse, and some photons are wasted. Hence, the total number of *PQ*s re-oxidized during a pulse and the ensuing $T_{dark}$ is given by

$$n_{re-oxidized} = \begin{cases} \dfrac{n_o T_{pulse}}{2} & , \quad T_{pulse} \le T_{pulse}^{max} \\ \dfrac{T_{pulse}}{\tau_{del}} + N_{pool} & , \quad T_{pulse} > T_{pulse}^{max} \end{cases} \qquad \text{(Equation 10)}$$

which yields an expression for $\eta_{ph}$ (from Equation 3):

$$\eta_{ph} = \begin{cases} (n_0/2)\,\tau_{del} & , \quad T_{pulse} \le T_{pulse}^{Max} \\ 1 \; + \; N_{pool}\left(\tau_{del}/T_{pulse}\right) & , \quad T_{pulse} > T_{pulse}^{Max} \end{cases}. \qquad \text{(Equation 11)}$$

## How $T_{pulse}$ Affects $\eta_{ph}$

Equation 11 shows the importance of selecting $T_{pulse}$ appropriately. For $T_{pulse} < T_{pulse}^{Max}$, $\eta_{ph}$ is constant. For $T_{pulse} > T_{pulse}^{Max}$, $\eta_{ph}$ diminishes as $T_{pulse}$ is increased, illustrated in Figure 5. For $A$ = 1 nm$^2$, a delivery time (bottleneck) of 10 ms, and $N_{Pool}$ = 7, one finds from Equations 6 and 9:

$$\begin{aligned} I = 1{,}000\ \mu E/(m^2 - s): \quad & n_0 = 0.602 \text{ photons/ms} \quad , \quad T_{pulse}^{max} = 34.81 \text{ ms} \\ I = 2{,}000\ \mu E/(m^2 - s): \quad & n_0 = 1.204 \text{ photons/ms} \quad , \quad T_{pulse}^{max} = 19.34 \text{ ms}. \end{aligned} \qquad \text{(Equation 12)}$$

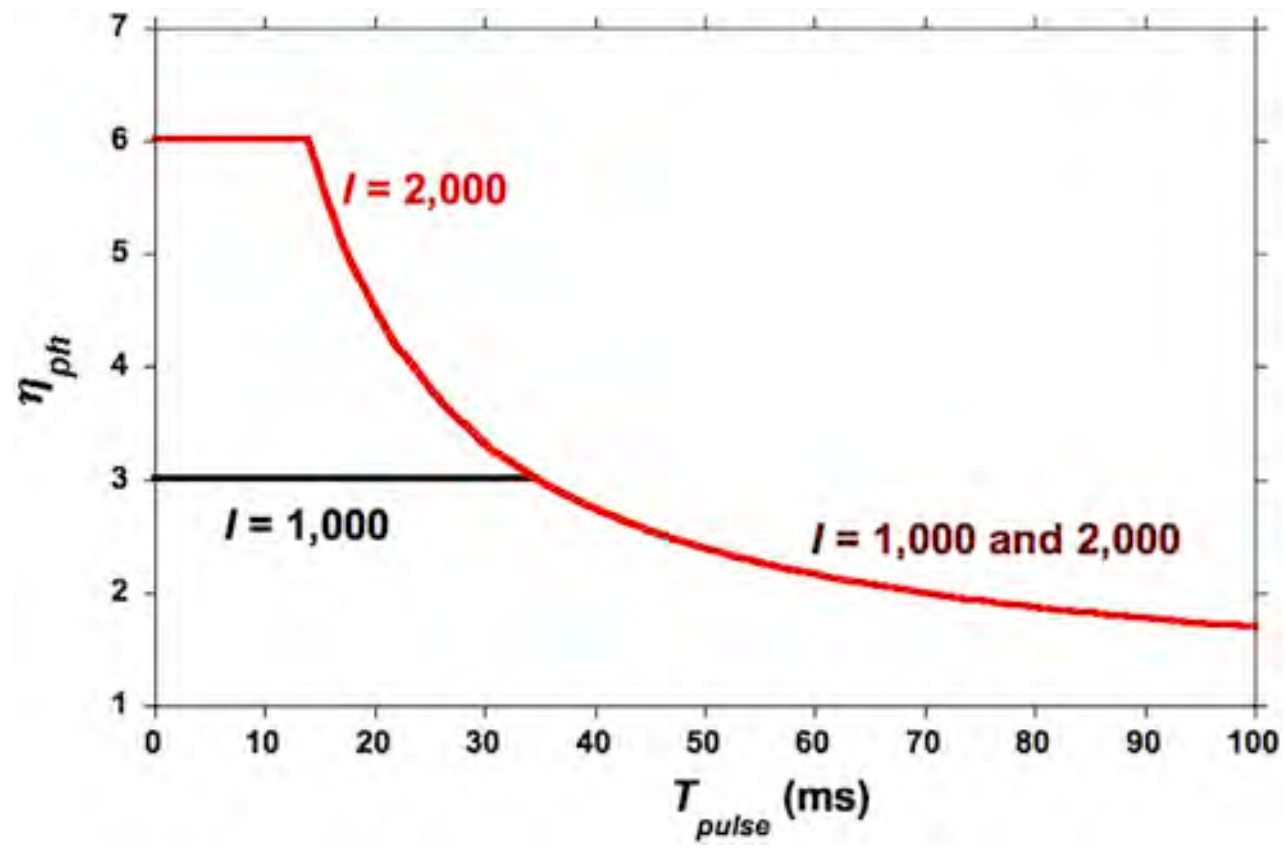

**Figure 5. Calculated $\eta_{ph}$ versus Pulse Duration**

At two values of photon flux density (from Equation 11). $A = 1\ nm^2$, $N_{Pool} = 7$, and $\tau_{del} = 10$ ms.

As long as $T_{pulse}$ does not exceed these maximal values, *all* reduced *PQ*s are re-oxidized during the dark period. Hence, using Equation 11:

$$\eta_{ph} = (n_0 / 2)\ \tau_{del}. \qquad \text{(Equation 13)}$$

The prediction is then $\eta_{ph} = 3$, which is the same value obtained in the new experimental results reported here, as well as in the data of Vejrazka et al. (2011), both at the same *I* value and both with pulse times not exceeding that of Equation 12.

## Estimating Effective Absorption Cross-Section *A*

Consider the results of Simionato et al. (2013) for $I = 1{,}200\ \mu E/(m^2\text{-}s)$. The observed *time-averaged* biomass production rates under continuous light and under a pulsed regime of $\{T_{pulse} = 10$ ms, $T_{dark} = 90$ ms$\}$ are close to one another. Using Equation 3, one finds that this implies $\eta_{ph} = 10$. In terms of Equation 11, this means that for every *PQ* re-oxidized under continuous light (roughly one every 10 ms), 10 *PQ*s are re-oxidized in one pulsed cycle. Based on Equation 7, roughly 1 *PQ* is re-oxidized during the pulse. To obtain $\eta_{ph}$ of order 10, the *PQ* pool must be able to store 9 *PQ*s. Equation 9 then yields

$$n_{additonal} = 9 = (n_0 / 2 \ - \ 1 / \tau_{del})\ T_{pulse}. \qquad \text{(Equation 14)}$$

For $T_{pulse} = \tau_{del} = 10$ ms, Equation 14 yields $n_0 = 2$ photons/ms. Equation 6 then requires $A = 2.77\ nm^2$. Changing $\tau_{del}$ within a reasonable range (5 ms $\leq \tau_{del} \leq$ 15 ms) does not significantly modify the resulting value of *A*.

## Comparison of Model Predictions against Our Measurements

Figure 6 summarizes comparisons between our measurements and the simple average model. For the case with $T_{pulse} = 150$ ms and $T_{dark} = 250$ ms, where the predicted trend is opposite to that of the data, we show in the Supplemental Information (Figures S1 and S2) how the trend can be predicted by the model if (1) photo-acclimation can cause a reduction of *A* as well as a lessening of $N_{pool}$ as *I* is increased (de Wijn and van Gorkom, 2001; Zou and Richmond, 2000; Simionato et al., 2011; Bonente et al., 2012; Gris et al., 2014) or (2) there is a distribution of $N_{pool}$ (Guemther et al., 1988; Hemelrijk and van Gorkom, 1996; Cleland, 1998).

## Comparison of Model Predictions against Data with Pulse Times ≥ 10 ms

The values for $\eta_{ph}$ deduced from the data of Simionato et al. (2013) are plotted in Figure 7. They correspond to a different algal strain and different photobioreactor conditions compared with the ones reported here. A larger *A* value was necessary for the model to yield good agreement ($A \approx 3.5\ nm^2$), but is within the range reported in the literature (de Wijn and van Gorkom, 2001; Zou and Richmond, 2000; Simionato et al., 2011; Bonente et al., 2012; Gris et al., 2014).

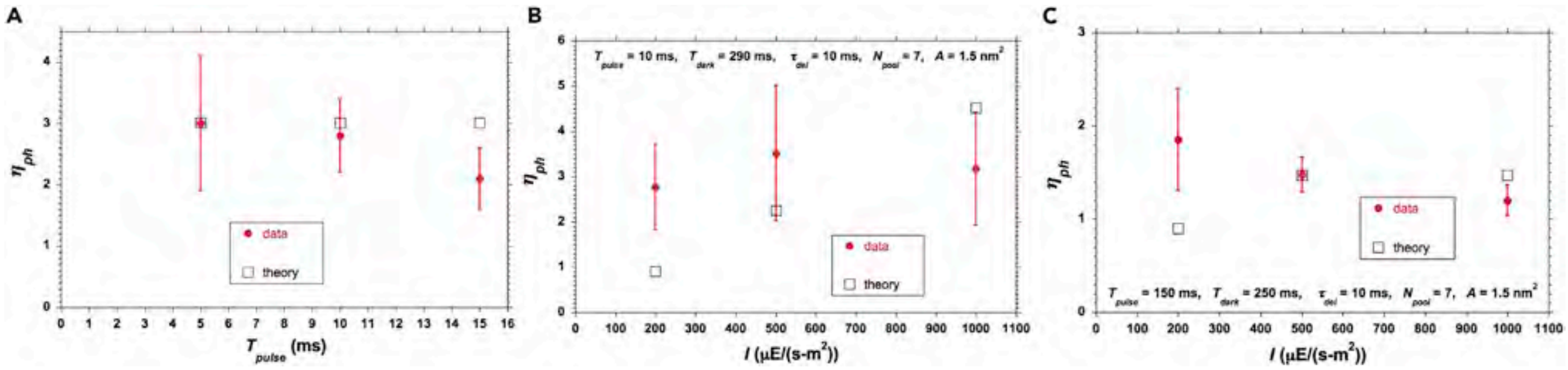

**Figure 6. Measured and Calculated $\eta_{ph}$**

$\eta_{ph}$ versus (A) $T_{pulse}$ at $I$ = 1,000 μE/(m$^2$-s) (at the optimal $T_{dark}$) and $N_{pool}$ = 7, $\tau_{del}$ = 10 ms, $A$ = 1 nm$^2$ for model calculations; (B) $I$ for $T_{pulse}$ = 10 ms, $T_{dark}$ = 290 ms; and (C) $I$ for $T_{pulse}$ = 150 ms, $T_{dark}$ = 250 ms. Vertical bars indicate ±1 standard deviation about the average for the experimental data, with each measured point coming from 8 replications.

## Comparison of Model Predictions against Data with Very Short Pulse and Cycle Times

The data reported in Vejrazka et al. (2011) were for a pulsed regime with $T_{pulse}$ = 1 ms, $T_{dark}$ = 9 ms, and $I$ = 1,000 μE/(m$^2$-s), for which a measured $\eta_{ph} \approx 3$ can be deduced. With $A$ = 1 nm$^2$, Equation 6 yields $n_0$ = 0.6 photons/ms, hence an average $PQ$ reduction rate of 0.3 molecules/ms. As all the $PQ$s are re-oxidized during a pulsed cycle, a delivery time of $\tau_{del}$ = 10 ms yields a computed value of $\eta_{ph} \approx 3$, in agreement with the data. However, the dynamics are more complicated because $T_{pulse}$ is much shorter than $\tau_{del}$, a point explored in the subsequent discussion.

## Uncertainties and Limitations

Values of $A$, $\tau_{del}$, and $N_{pool}$ were not reported in the previous studies from which $\eta_{ph}$ could be estimated from their data (Vejrazka et al., 2011, 2012, 2013, 2015; Simionato et al., 2013). Moreover, it remains to be established whether such parameters depend on photo-acclimation. This is important because, in our experiments, algae were photo-acclimated to each separate pulsed regime. Nevertheless, the non-negligible variations of the values of these parameters can be appraised from several prior studies (de Wijn and van Gorkom, 2001; Zou and Richmond, 2000; Simionato et al., 2011; Bonente et al., 2012; Gris et al., 2014; Ooms et al., 2016) and typically vary from one determination to another by no more than a factor of 2–3 (as opposed to an order of magnitude). The more important variances are in other phenomena and parameters; hence we do not attempt to find best-fit values for this parameter set. Rather, we try to demonstrate that the simple picture of rate-limiting behavior depicted in Figure 3, combined with the average photon arrival times, can account for a wide range of measurements of $\eta_{ph}$ in pulsed versus continuous irradiation experiments.

The simple average model does not take into account that the numbers of $PQ$s reduced and stored are integers. Hence, the model will generate an error that may be large when the number of $PQ$s reduced per pulse is small, which can stem from very short pulses, low $I$, small $A$, and/or small $N_{pool}$. Second, this

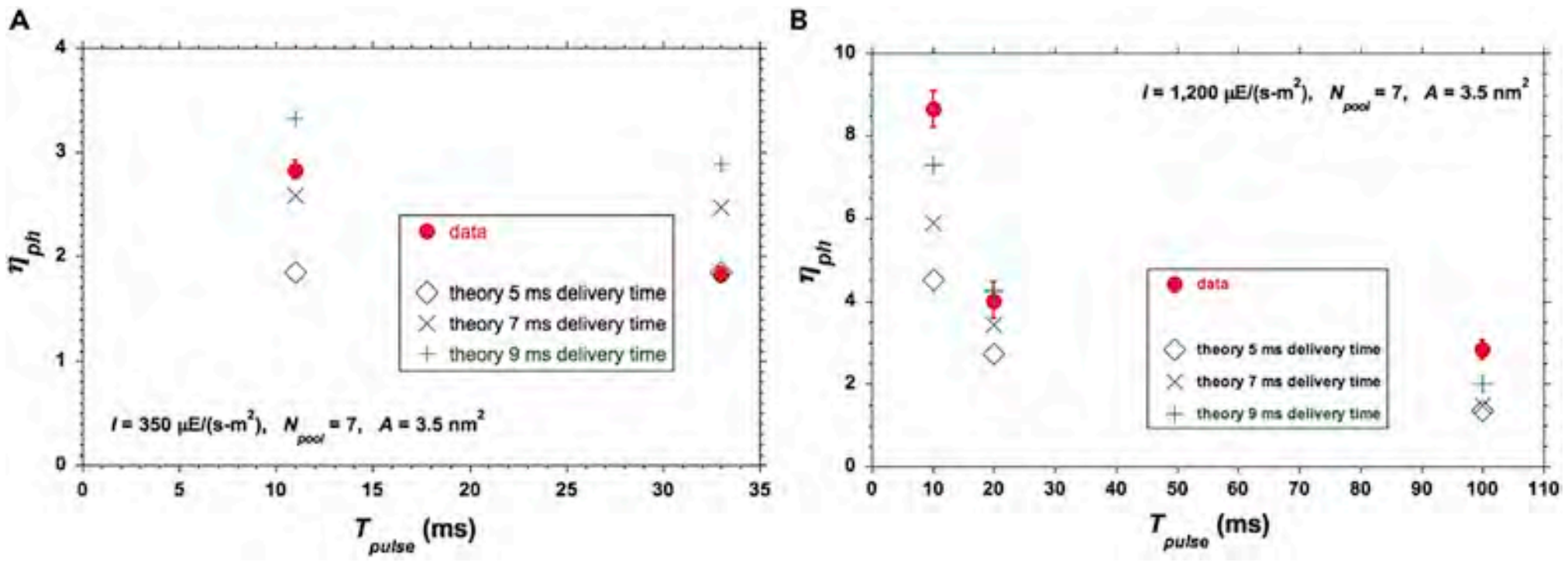

**Figure 7. Comparison of Simple Average Model Predictions for $\eta_{ph}$ against Data with Pulse Times ≥ 10 ms**

Calculated values show the sensitivity to the assumed $\tau_{del}$. (A) For two pulse durations at a relatively low $I$ = 350 μE/(m$^2$-s). (B) For three pulse durations at an intermediate $I$ = 1,200 μE/(m$^2$-s). The vertical bars of ±1 standard deviation about the average of each measured data point were taken from the original reference Simionato et al. (2013) with no further elaboration therein.

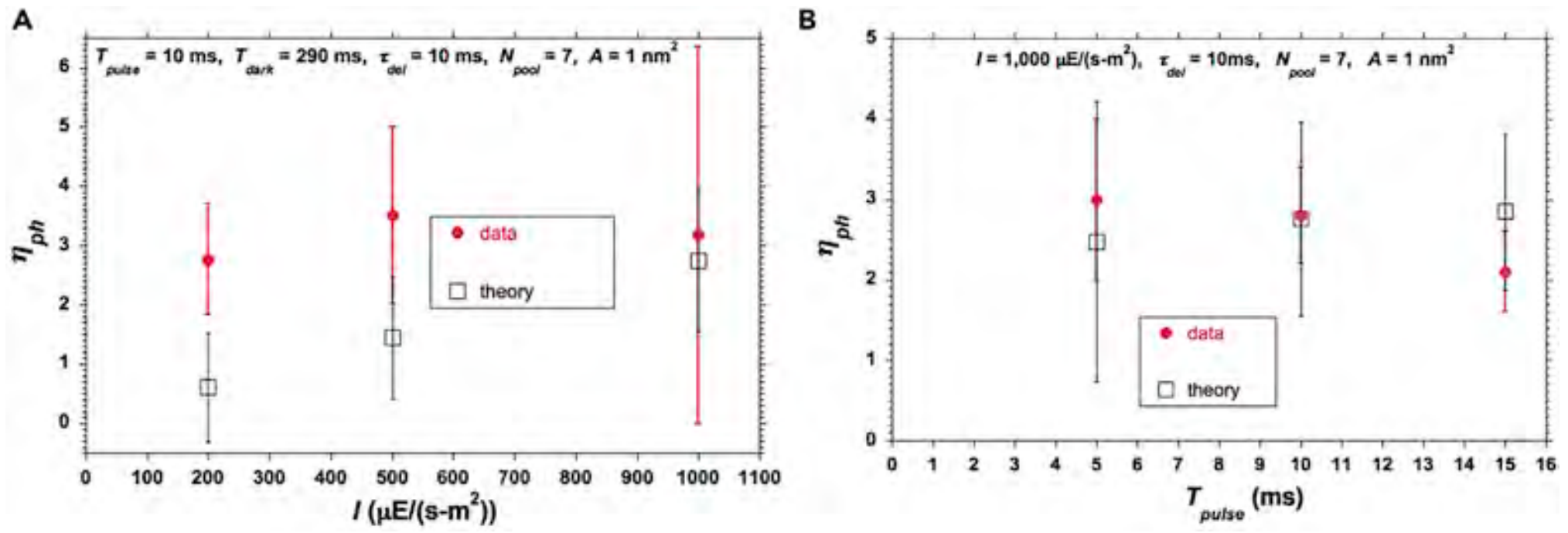

**Figure 8. Comparisons of Model Predictions Against Data for $\eta_{ph}$**

For (A) $I$ = 200, 500, and 1,000 $\mu$E/($m^2$-s), at $T_{pulse}$ = 10 ms and $T_{dark}$ = 290 ms; (B) $T_{pulse}$ = 5, 10, and 15 ms, at $I$ = 1,000 $\mu$E/($m^2$-s). The vertical bars of $\pm 1$ standard deviation about the average (1) come from 8 replications for each measured point and (2) correspond to the inherent standard deviations associated with photon arrival statistics for the model (computed) results, as elaborated in the text.

version of the model cannot account for the effect of photon arrival-time statistics or the existence of a *PQ* pool size distribution on the statistical fluctuations in *PQ* reduction rates, and hence cannot provide an estimate of standard deviations. These limitations call for a full statistical analysis.

## Full Statistical Analysis

### *Model Details*

A random sequence of photon arrivals times was prepared. The program then counted how many *PQ*s are reduced from pairs of photons during a pulse, how many of them are re-oxidized to deliver charges to the next stage of the production process during the pulse (*n*(*reoxidized*)), and how many remain reduced in the *PQ* pool.

The second in a pair of consecutive photons is lost if the time between photon arrivals is smaller than 0.2 ms. Both photons are lost if all the *PQ*s are already reduced. At the end of the pulse, the program generates:

1 The number of reduced *PQ*s in the pool: n(kept).

2 The number of *PQ*s re-oxidized during a full cycle of duration $T_{pulse} + T_{dark}$:

$$\mathrm{n(reoxidized)} = \mathrm{n(reoxidized\ during\ flash)} + \mathrm{n(kept)} \qquad \text{(Equation 15)}$$

3 The number of photons lost during a pulse: n(lost).

Similar runs for very long times generated the same quantities for continuous light. The output yields the average rate of *PQ* re-oxidation, $<r(re\text{-}oxidized)>_{Continuous}$. For a given light intensity and parameter set ($I$, $A$, $\tau_{del}$, $N_{Pool}$), the average $\eta_{ph}$ was computed as:

$$\eta_{ph} = \frac{\langle \mathrm{n(reoxidized)}_{flash} \rangle / \mathrm{T}_{flash}}{\langle \mathrm{r(reoxidized)}_{continuous} \rangle}. \qquad \text{(Equation 16)}$$

The program also computed the standard deviation around this average.

In addition, the program generated the probability distributions of *n*(*kept*), *n*(*re-oxidized*), and *n*(*lost*). Representative results are presented in the Supplemental Information (Figure S3). To check the sensitivity of the statistical results to sample size, computations were performed over $10^5$, $10^6$, and $10^7$ reaction centers. The program generated the average and standard deviation of the desired quantities (Supplemental Information). The values obtained with different sample sizes did not vary in a statistically significant manner. Hence, we computed all results for a sample of $10^6$ reaction centers.

## Comparisons between Predictions of the Full Statistical Model against Our Data

Our data and the corresponding model predictions in Figure 8 highlight the sensitivity of $\eta_{ph}$ to $I$ and $T_{pulse}$. The standard deviations noted for the *calculated* results derive from the inherent variance

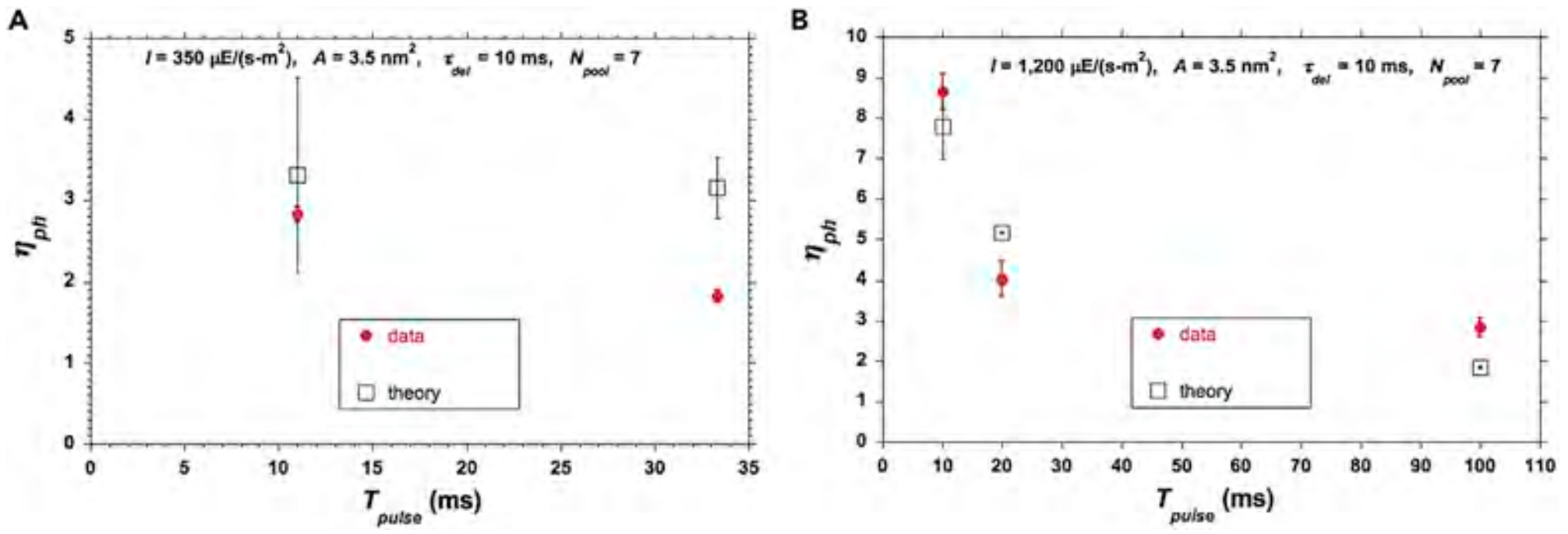

**Figure 9. Comparisons of Model Predictions against Data for $\eta_{ph}$**

For (A) ($T_{pulse}$, $T_{dark}$) = (11 ms, 22 ms), and ($T_{pulse}$, $T_{dark}$) = (33.33 ms, 66.67 ms), at a relatively low value of $I$ = 350 $\mu$E/(m$^2$-s); (B) ($T_{pulse}$ $T_{dark}$) = (10 ms, 20 ms), ($T_{pulse}$, $T_{dark}$) = (20 ms, 180 ms), and ($T_{pulse}$, $T_{dark}$) = (100 ms, 900 ms) at $I$ = 1,200 $\mu$E/(m$^2$-s). The vertical bars of $\pm$1 standard deviation about the average (1) were taken from Simionato et al. (2013) with no further elaboration therein and (2) correspond to the inherent standard deviations associated with photon arrival statistics for the model (computed) results, as elaborated in the text.

associated with photon arrival-time statistics. For the relatively long $T_{pulse}$ = 150 ms, the predicted trend is opposite to that of the data even when the effect of photon arrival-time statistics is accounted for. Different choices of model parameters could reduce the discrepancy, but could not eliminate it. In the Supplemental Information (Figure S4), we show that the discrepancy can be remedied by taking into account the possibility that $A$ and $N_{pool}$ may vary owing to photo-acclimation or to the existence of a distribution of $PQ$ pool size.

## Comparisons between Predictions of the Full Statistical Model against Data with Pulse Times $\geq$ 10 ms

An additional comparison, based on the experimental results from Simionato et al. (2013), is offered in Figure 9 where a larger but reasonable value of $A$ was again necessary to achieve reasonable agreement.

## Analysis with Data with Ultra-Short Pulses and Dark Times

Values of $T_{pulse}$ shorter than $\tau_{del}$ pose an intriguing challenge, because there are situations where, on average, not even a single photon impinges upon $A$ within the pulse duration. Of particular interest are the data from Vejrazka et al. (2011) with $T_{pulse}$ = 1 ms and $T_{dark}$ = 9 ms (at $I$ = 1,000 $\mu$E/(m$^2$-s)), for which one can deduce that $\eta_{ph} \approx 3$. Photon arrival-time statistics plays a crucial role here owing to the shortness of the pulse. Furthermore, $T_{dark}$ = 9 ms is substantially shorter than the dark times employed in all other experiments for which adequate data were available to perform the analyses ($T_{dark}$ = 200–300 ms in our measurements, and $T_{dark}$ = 20–900 ms in those of Simionato et al., 2013). It will now be shown that convolving the shortness of $T_{pulse}$ (1 ms) with photon arrival-time statistics leads to an *effective* $T_{dark}$ that may be substantially longer than the nominal 9 ms.

For $I$ = 1,000 $\mu$E/(m$^2$-s) and $A$ = 1 nm$^2$, the average arrival time between consecutive photons is 1.66 ms. This corresponds to a probability of 0.45 for one photon arriving during a pulse. The probability of receiving two photons in two consecutive pulses is then 0.205. This means that, typically, two photons will be absorbed by cross-section $A$ in two consecutive pulses only once every 5 rounds of two pulses, namely, every 100 ms.

Another possible scenario is cross-section $A$ absorbing two photons at least 0.2 ms apart in one pulse, with 0.2 ms being the shortest rate-limiting timescale of interest. The probability for such an event is 0.075. Hence, it occurs, on average, every 13 cycles, amounting to an effective $T_{dark}$ > 100 ms. Using the detailed statistical analysis to compute the average number of $PQ$s reduced per photon, we find that events solely of this kind yield $\eta_{ph}$ = 2.52, with a large standard deviation. Within the statistical error bars, this is consistent with the data.

In the more probable scenario of only one photon being absorbed during a pulse, a singly reduced $PQ$ molecule ($QA^-$) is generated. Its lifetime determines the probability of its surviving the long *effective* dark time, of the order of 100 ms, so as to be affected by a second photon, completing the generation of the doubly reduced *PQ* molecule ($QB^{-2}$) required for biomass production. In view of the fact that the *effective* $T_{dark}$ is an order of magnitude longer than the nominal dark time of 9 ms, the data analyzed here appear to indicate that the lifetimes of $QA^{-}$ and $QB^{-2}$ must be long enough to survive these long effective dark times to which absorption cross-section *A* is exposed.

The short 1-ms pulse also points to the importance of photon arrival statistics. Consider a longer $T_{pulse}$, e.g., 5 ms. The probability of $A = 1\ nm^2$ receiving a single photon is then 0.95, and the probability of receiving two photons at least 0.2 ms apart is 0.70. Therefore, most reaction centers *would* receive two photons in a single pulse under this parameter set.

## Effects of *PQ* Pool Size Distribution and Photo-acclimation

Information about the existence of a distribution of *PQ* pool sizes is scant. For the data published to date (Hemelrijk and van Gorkom, 1996; Cleland, 1998), we found that a normal probability distribution provides a good fit:

$$P(N_{Pool}) = \frac{1}{\sqrt{2\pi\sigma^2}}\, e^{-\frac{(N_{Pool} - \langle N_{Pool}\rangle)^2}{2\sigma^2}} \quad \text{(Equation 17)}$$

where $\langle N_{Pool}\rangle$ is the average pool size and $\sigma$ is the standard deviation. A fit to the results of Hemelrijk and van Gorkom (1996) yields $\langle N_{Pool}\rangle = 6–7$ with $\sigma = 1–3$.

For the relatively long $T_{pulse} = 150$ ms, the data (Supplemental Information) show $\eta_{ph}$ decreasing with *I*, whereas the model predicts the opposite trend. However, the model assumed a *fixed PQ* pool size. For such long pulses, $\eta_{ph}$ becomes particularly sensitive to $N_{Pool}$, because the number of *PQ*s reduced per pulse is large. The ability to store most or all of the reduced *PQ*s depends on whether $N_{Pool}$ varies with *I* and/or whether there is a distribution of pool sizes. The capability of the model to account for the correct behavior is presented in the Supplemental Information (Figures S5 and S6).

# DISCUSSION

## Translating Higher Photon Efficiency to Increased Bioproductivity

The enhanced $\eta_{ph}$ in pulsed-light experiments comes at the price of low time-averaged bioproductivity. Designing photobioreactors for ultra-high bioproductivity is challenging, but solutions *are* possible, e.g., opto-mechanically manipulating the distribution of light input such that delivered photons are not wasted while each reactor is exposed to the requisite light-dark cycles.

Another direction is inducing suitably turbulent mixing in dense cultures under *continuous* irradiation whereby *effective* light-dark cycles are experienced by algal cells. This was achieved in Qiang et al. (1998a, 1998b) and Richmond et al. (2003), where turbulent mixing was induced by gas bubbles fed at the bottom of a thin vertical channel illuminated on both sides with *continuous* halogen-lamp light at $I = 250–4{,}000\ \mu E/(m^2\text{-}s)$. Culture densities were so high that the photic zone was only ~1 mm (Qiang et al., 1998a, 1998b; Richmond et al., 2003). Fluid turbulence induced random cell motion with an effective diffusion coefficient of order 1 $cm^2/s$, ensuring that the average time spent by cells in the photic zone was of order ~0.2–2 ms (Gebremariam and Zarmi, 2012; Greenwald et al., 2012; Zarmi et al., 2013). Each short *effective* light pulse delivered a small number of photons to cross-section *A*. Hence, despite nominally continuous irradiation, "clogging" in PS II could be avoided. The cells then spent ~200–400 ms in the dark region, enough time to process the absorbed photon energy, during which other cells migrated into the photic zone.

Saturation of the *P-I* curve was avoided, while photon efficiency was raised, with no signs of photo-inhibition. The paucity of any evidence of photo-inhibition at such high light intensities accentuates the fact that photo-inhibition is determined by *cumulative* photon absorption, which can be maintained sufficiently small by applying intense light pulses for only a small fraction of the cycle time, be it with properly pulsed light-emitting diodes or via suitable turbulent mixing of the algal culture. In these experiments, *P* exceeded the rates obtained in standard reactors by a factor of ~3 (Qiang et al., 1998a, 1998b; Richmond et al., 2003). Furthermore, as *I* was increased to 4,000 $\mu E/(m^2\text{-}s)$, *P* grew almost linearly with *I*, with an *absolute* (time-averaged) photon efficiency of 15% (based on photosynthetically active radiation), which is close to the thermodynamic limit (Gordon and Polle, 2007).

We have presented data from pulsed-light experiments on algal photobioreactors, complemented by simple physical arguments rooted in photon arrival-time statistics, to substantiate significant increases in the relative photon efficiency of algal photosynthesis. The key is identifying the principal rate-limiting photosynthetic step in PS II, and imposing a judiciously chosen pulsed-light regime for a given photon flux density, so as to attain the necessary synchronization of biological and photonic timescales. The enhancement in relative photon efficiency varies from a factor of 3 (from our own measurements and deduced from prior studies) to a factor of 10 (deduced from published data).

This enhancement does not automatically enable the practical attainment of higher bioproductivity in a scalable cultivation device, for which skillful optical, mechanical, and hydrodynamic design of photobioreactors is required. Indeed, previous investigations (Qiang et al., 1998a, 1998b; Richmond et al., 2003) realized the commensurate improvement in bioproductivity via a combination of turbulent mixing, dense cultures, and thin reactors. The challenge of engineering feasible photobioreactors that can achieve this objective, be they driven by solar or artificial light, is delegated to future research efforts. (The use of artificial light should be viewed as sustainable provided the electricity source derives from renewables such as solar, wind, or hydroelectric.)

We have identified the associated timescales, as well as an understanding of how synchronization between the pulsed-light regime and biological timescales can lead to markedly enhanced photon efficiency.

## Limitations of the Study

Our experimental and modeling results prompt fundamental questions in algal research, for which experimental results are needed before properly optimized reactors can be designed. These issues subsume:

Photo-acclimation: How does algal performance depend on acclimation to pulsed-light regimes (in particular on millisecond timescales, including the dependence on instantaneous photon flux density)? How does $A$, as well as the size and distribution of $N_{pool}$, vary with these pulsed regimes and with light intensity?

Optimal dark time: The basis for quantifying the optimal dark time under pulsed light is not yet understood and requires detailed study. Increasing bioproductivity by following each pulse with a sufficiently long dark time was proposed (Abu-Ghosh et al., 2015), but the dark times in that study were too short for the dramatic potential improvement in photon efficiency depicted here to be observed.

Long timescales characterizing post-PS II processes: These long timescales do not appear to affect the *P-I* curve. Is it because these processes are endowed with large buffers for storing intermediate products or is it because there are parallel-processing elements in ensuing stages?

*PQ* pools: Their size and possible size distribution need to be ascertained.

Genetic intervention: To what extent can it further improve photon efficiency via modification of effective absorption cross-section $A$ and *PQ* pool size? For example, our model predicts that increasing $A$ can increase $\eta_{ph}$ proportionately for commensurately modified pulse regimes, with genetic intervention already having demonstrated the ability to moderate $A$ (Melis et al., 1998, 1999; Polle et al., 2002, 2003; Kirst and Melis, 2014), whereas increasing *PQ* pool capacity should not impact $\eta_{ph}$ but would affect the tolerance to high light intensity and would allow for a wider range of pulse-time duration.

## METHODS

All methods can be found in the accompanying Transparent Methods supplemental file.

## SUPPLEMENTAL INFORMATION

Supplemental Information can be found online at https://doi.org/10.1016/j.isci.2020.101115.

## ACKNOWLEDGMENTS

Y.Z. and J.M.G. express appreciation for financial support for this research from Reliance Industries Ltd. J.M.G. gratefully acknowledges the hospitality and support of the Institute of Advanced Studies of the University of Western Australia during the writing of this paper. The RIL authors gratefully acknowledge Sridharan Govindachary for his technical input about the photosynthetic process.

## AUTHOR CONTRIBUTIONS

Y.Z.: conceptualization, investigation, visualization, methodology, validation, formal analysis, writing, review & editing; J.M.G.: conceptualization, investigation, visualization, methodology, validation, formal analysis, writing, review & editing; A.M.: investigation, methodology, data curation, formal analysis, writing, review & editing; A.R.K.: conceptualization, validation, resources, supervision, writing, review & editing; S.D.P.: investigation, data curation, methodology, writing, review & editing; A.B.: conceptualization, validation, writing, review & editing; B.G.R.: investigation; T.P.G.: conceptualization, project administration, writing, review & editing; A.S.: conceptualization, funding acquisition, writing, review & editing.

## DECLARATION OF INTERESTS

The authors have no conflict of interest to declare.

## Supplemental Information

# Enhanced Algal Photosynthetic Photon Efficiency by Pulsed Light

**Yair Zarmi, Jeffrey M. Gordon, Amit Mahulkar, Avinash R. Khopkar, Smita D. Patil, Arun Banerjee, Badari Gade Reddy, Thomas P. Griffin, and Ajit Sapre**

# Supplemental Information

<u>Results when $N_{pool}$ and $A$ can vary</u>

For the case with $T_{pulse}$ = 150 ms and $T_{dark}$ = 250 ms (Fig. S1a), the predicted trend is opposite to that of the data. Whereas the measured $\eta_{ph}$ decreases as $I$ is increased, the model predicts an increase in $\eta_{ph}$ (Eq. (11)). The predicted increase is predicated on both $A$ and $N_{pool}$ having constant values. However, photo-acclimation can cause a reduction in $A$ and $N_{pool}$ as $I$ is increased (de Wijn and van Gorkom, 2001; Zou and Richmond, 2000; Simionato et al., 2011; Bonente et al., 2012; Gris et al., 2014). Figure S1b illustrates that the correct trend can be attained if it is assumed that photo-acclimation induces changes in $A$ and $N_{Pool}$ as $I$ is increased (Table S1 lists values that are reasonable based on prior measurements, but do not signify actual observed parameters).

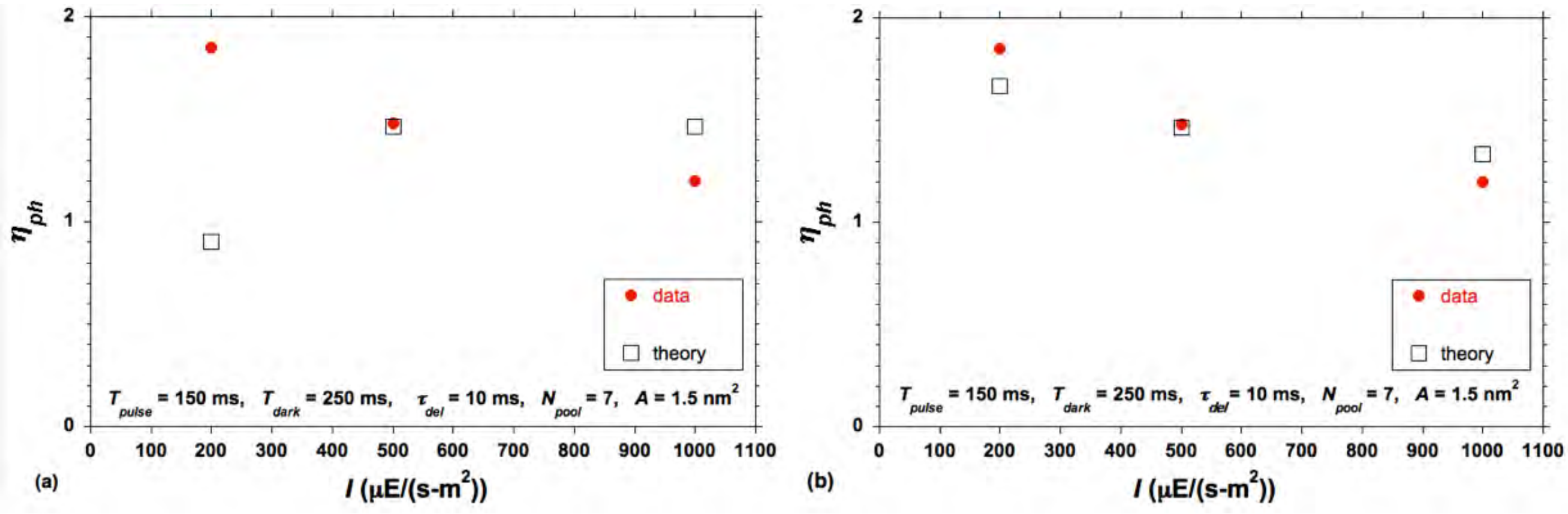

<u>Fig. S1</u>. Related to Fig. 6c. (a) Measured and calculated (Eq. (11)) $\eta_{ph}$ vs. $I$ for $T_{pulse}$ = 150 ms, $T_{dark}$ = 250 ms (same as Fig. 6c where the standard deviations and number of replications for the measured points are noted). (b) Modification of model predictions of part (a) when model parameters vary with $I$ owing to photo-acclimation as in Table S1. The data points are the same as in part (a).

Figure S2a shows model predictions when a distribution of $N_{pool}$ is accounted for (Eq. (17)), with average $\langle N_{pool} \rangle$ = 7 and standard deviation $\sigma$ = 2. Figure S2b shows the same data but with model calculations that allow for the effect of photo-acclimation on $A$ and $N_{pool}$ as listed in Table S2. The effect of photon arrival time statistics has been incorporated.

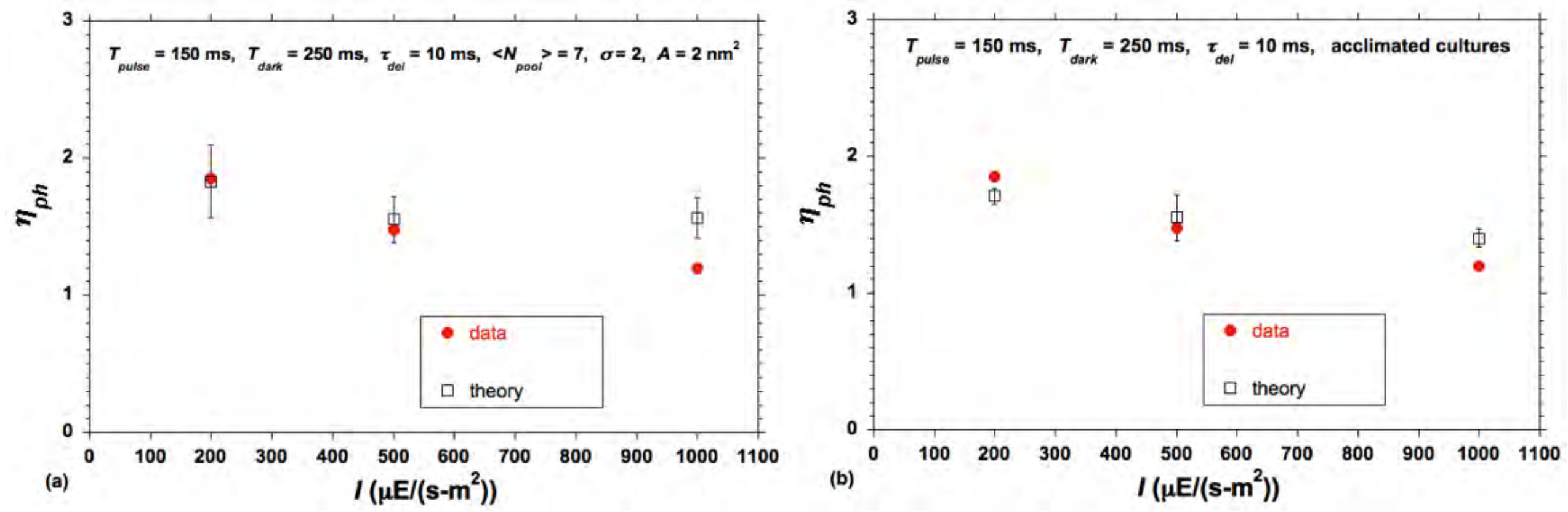

Fig. S2. Related to Fig. 6c. Dependence of $\eta_{ph}$ on $I$ at the relatively long $T_{pulse}$ = 150 ms. The measured data and their standard deviations are the same as in Fig. S1. Model predictions account for (i) a distribution of $N_{pool}$, and (ii) the possible impact of photo-acclimation on $A$ and $N_{pool}$. Vertical bars for the model (computed) results correspond to the inherent standard deviations associated with photon-arrival statistics, as elaborated in the text.

Figure S3 further sharpens this point with a comparison between data from (Simionato et al., 2013) and model predictions when a distribution of $N_{pool}$ is accounted for. In all these computations with the full statistical model, each reaction center was randomly assigned a value of $N_{pool}$ using the probability density of Eq. (17).

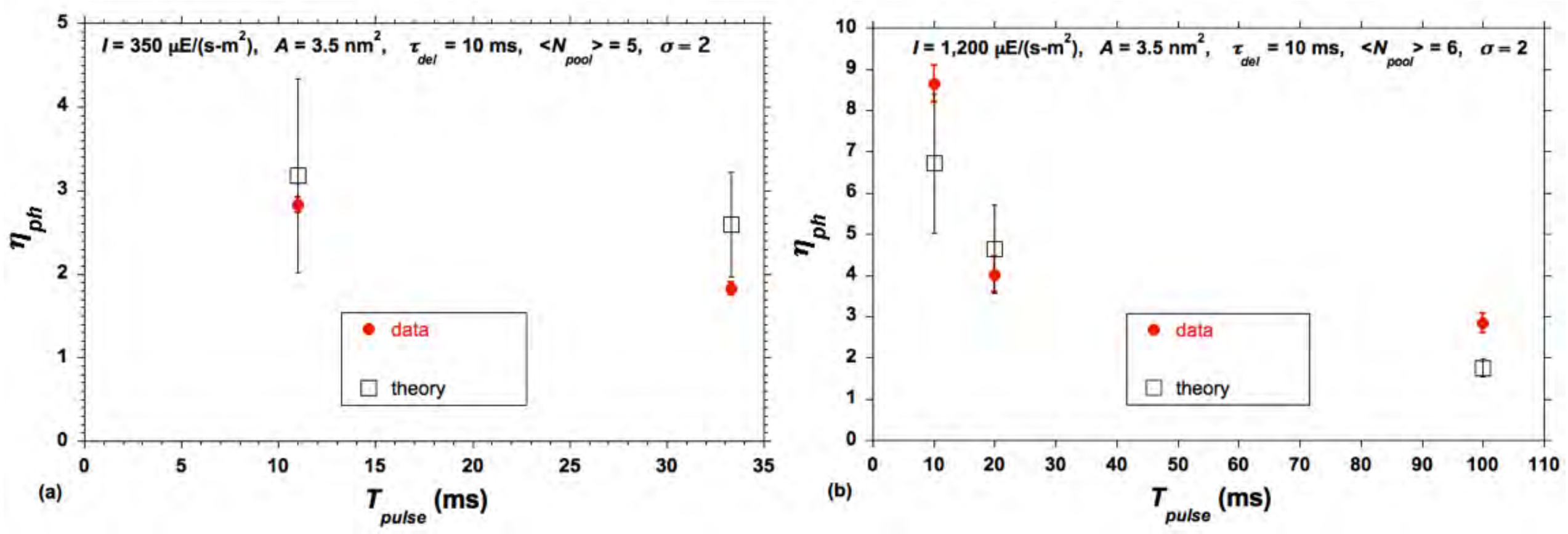

Fig. S3. Related to Fig. 7. Comparison of data from (Simionato et al., 2013) against model predictions for the dependence of $\eta_{ph}$ on $T_{pulse}$ at (a) low and (b) intermediate $I$ values. The theory accounts for a distribution of $N_{pool}$. The vertical bars of ±1 standard deviation about the average: (i) were taken from (Simionato et al., 2013) for the measured data, for 3 replications, and (ii) correspond to the inherent standard deviations associated with photon-arrival statistics for the model (computed) results, as elaborated in the text.

Table S1. Related to Fig. 6c. Parameter values used for the model predictions in Fig. S1, based on the possible impact of photo-acclimation.

| $I$ ($\mu$E/(m$^2$-s) | $A$ (nm$^2$) | $N_{Pool}$ |
|---|---|---|
| 200 | 4 | 10 |
| 500 | 2 | 7 |
| 1000 | 1 | 5 |

Table S2. Related to Fig. 6c. Parameter values for model calculations in Fig. S2, accounting for the possible effect of photo-acclimation.

| $I$ ($\mu$E/(m$^2$-s)) | $A$ (nm$^2$) | $N_{Pool}$ |
|---|---|---|
| 200 | 4 | 9 |
| 500 | 2 | 7 |
| 1000 | 1 | 5 |

Full statistical analysis

Sample distributions are presented in Fig. S4 for $I$ = 1000 $\mu$E/(m$^2$-s), $A$ = 1 nm$^2$, $T_{pulse}$ = 10 ms, $\tau_{del}$ = 10 ms and $N_{Pool}$ = 7. Figure S4a reflects the fact that the number of photons absorbed during the pulse is small. The average number of photons hitting $A$ = 1 nm$^2$ during a 10 ms pulse is 6.02. The actual number varies randomly from one reaction center to another, depending on the randomly varying time gaps between photons. Hence, the probability of fully reducing the $PQ$ pool is negligible, as is the probability of losing a large number of photons.

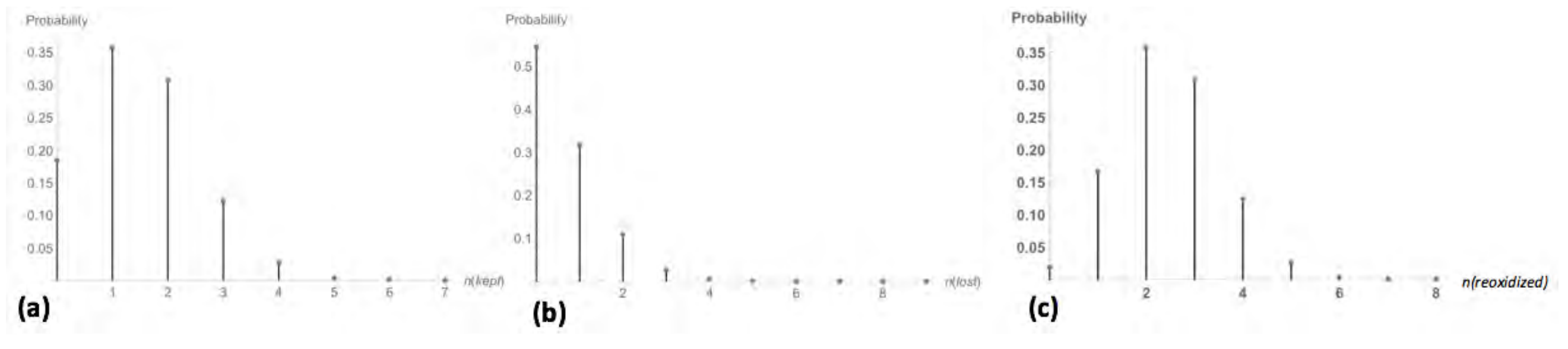

Fig. S4. Related to Fig. 8. Probability distribution for the number of (a) $PQ$s stored in the pool, (b) photons lost by the end of a 10 ms pulse, (c) $PQ$s reduced over a cycle with $T_{pulse}$ = 10 ms and a sufficiently long dark time.

The situation is quite different for long pulses. Examples are presented in Figs. S5 and S6 for $I$ = 1,000 μE/(m$^2$-s), $A$ = 1 nm$^2$, $\tau_{del}$ = 10 ms and $N_{Pool}$ = 7, but with a longer $T_{pulse}$ = 150 ms. Owing to the long pulse time, the probability that the *PQ* pool is completely reduced at the end of the pulse is high (Fig. S5a). The high number of lost photons (Fig. S6) is a consequence of the fact that the *PQ* pool is fully reduced shortly after the start of the pulse, so that many photons are lost due to reduced charge carriers.

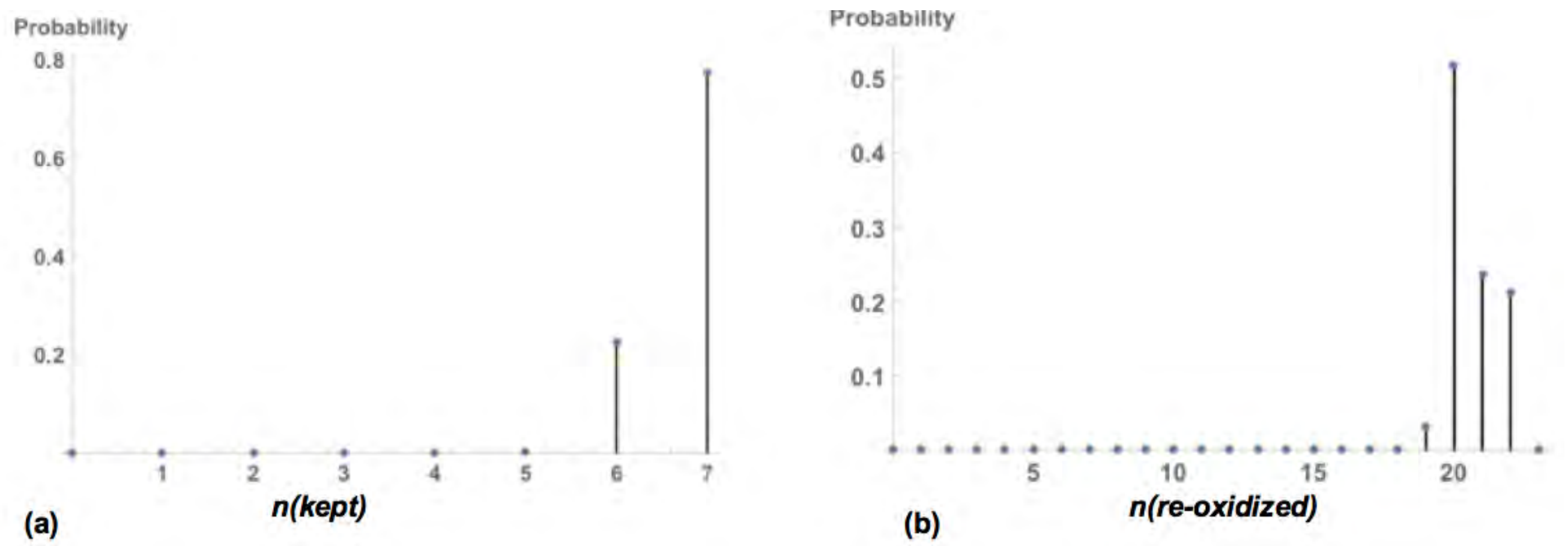

Fig. S5. Related to Fig. 6c. Probability distribution for (a) the number of *PQ*s reduced in the pool at the end of a 150 ms pulse and (b) the number of *PQ*s re-oxidized by the end of a cycle with $T_{pulse}$ = 150 ms and a sufficiently long dark time.

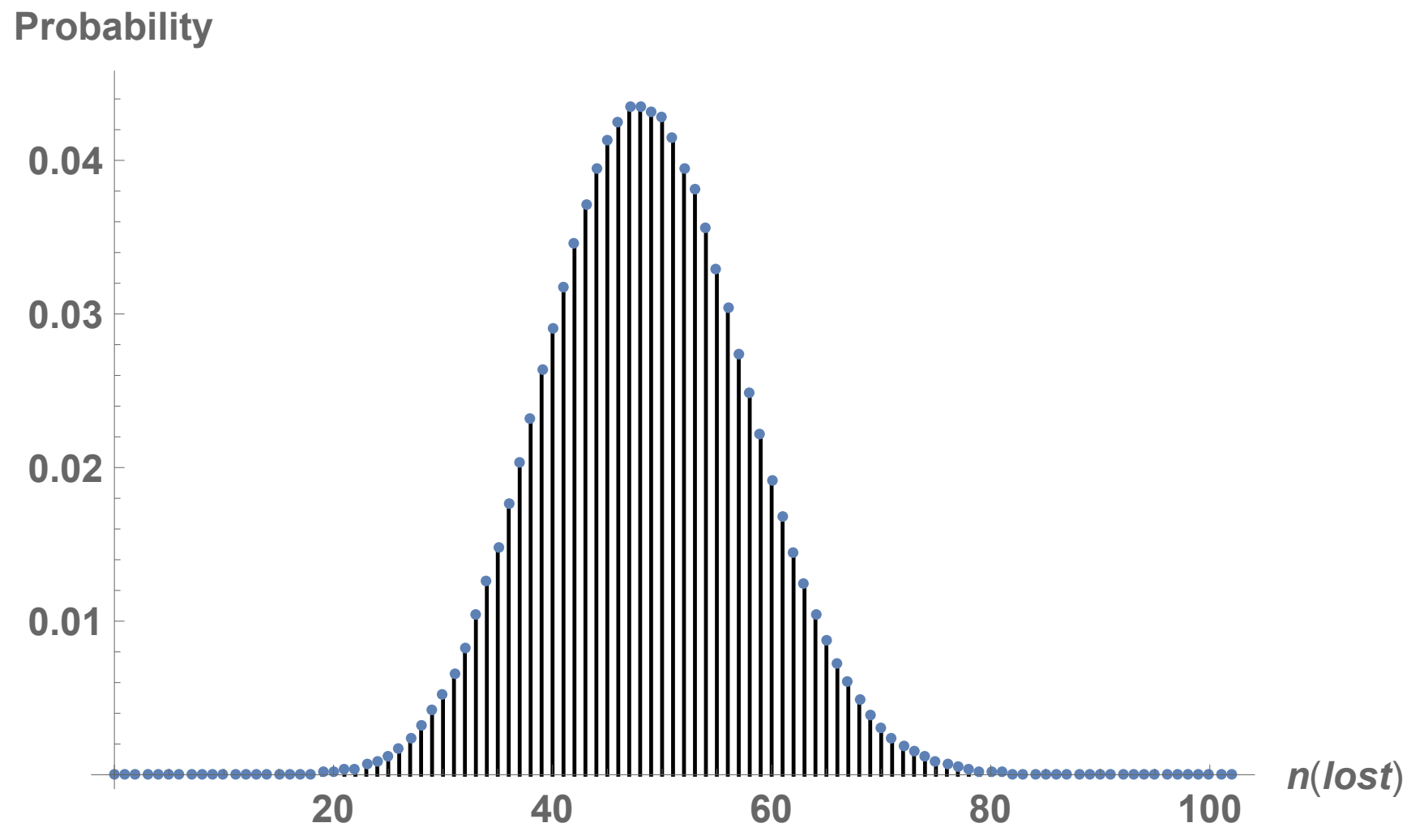

Fig. S6. Related to Fig. 6c. Probability distribution for the number of photons lost during a 150 ms pulse.

Using the approach of the simple average model, Eq. (5) yields the number of reduced *PQ*s to be 45. Of these, 15 are re-oxidized during the pulse (Eq. (7)), from which 7 *PQ*s can stay reduced. Hence, 22 *PQ*s can be re-oxidized during one cycle. The statistics of photon arrival times, combined with the losses owing to the 0.2 ms time scale, reduce this number to an average of 20.63, in which case $\eta_{ph} = 20.63/(T_{pulse} \cdot \tau_{del}) = 20.63/(150 \cdot 10) = 1.375$.

## Transparent Methods

We used a locally isolated strain of the *Nanochloris* species (which is a green microalgae) from our repository, cultivated in urea-phosphoric acid medium (urea 214 ppm and phosphoric acid 31 ppm, prepared in artificial seawater 4% by weight) and maintained at 27°C and 300 μE/(m$^2$-s), at an optical density of 2 in a 500 ml flask.

Biomass growth curves were generated using a Multi-Cultivator MC 1000-OD of Photon Systems Instruments (PSI, Czech Republic), comprising 8 test-tubes, each holding 70 ml of algal culture, and immersed in a water bath maintained at 35°C. Dilute algal cultures were used (density 17-30 mg/l, i.e., 0.05 OD, < 10% light attenuation), to ensure that all cells experienced essentially the same light intensity. pH was maintained at 7.0 by sparging humidifed air with 2% $CO_2$ at an air flow rate of 1 VVM (70 ml/min). Each test tube was irradiated by its own cool-white LED array.

For pulsed-light experiments, four PSI light sources (Model SL-3500, with an LC-100 PSI light controller) permitted independently tuning the irradiation and dark times from 1 ms to 999 ms. Our small glass reactors had optical path 3 cm, width 10 cm and height 15 cm. An operating height of 10 cm was used, so the total fluid volume was 300 ml.

For both continuous and pulsed irradiation, LEDs were controlled such that the instantaneous photon flux density for photosynthetically-active radiation at the surface of the culture was 1000 μE/(m$^2$-s), measured using Apogee Instruments' quantum meter model MQ-200. The 13 cm × 13 cm LED panel was sited less than 1 cm from the reactor. Light intensity was measured at the center of each of 9 equal-area regions comprising the reactor's illuminated surface, and the reported $I$ = 1,000 μE/(m$^2$-s) represents the average over these 9 sections.

Growth was measured over an illumination period of 6 hours, followed by a period of 6 hours of dark time, after which a fresh run was started, with these cycles repeated for 24 hours. Each day, the culture was harvested and brought to the desired starting operating optical density of 0.08, measured at a wavelength of 750 nm at the start ($OD_1$) and end of irradiation ($OD_2$) to get specific growth rate $\mu$ over time $t$: $\mu = \frac{ln(OD_2/OD_1)}{t}$. All runs were repeated ~20 times, from which average and standard deviation values were determined.